\documentclass[useAMS,usenatbib]{mn2e} %Original
\usepackage[dvips]{graphicx}
\usepackage{longtable}
\def \aj {AJ}
\def \apj {ApJ}
\def \apjl {ApJL}
\def \nat {Nature}
\def \mnras {MNRAS}
\def \aap {A\&A}
\def \araa {ARA\&A}
\def \pasp {PASP}

\begin{document}
\title[GRB 050401 : Observations and Modeling]{Optical Observations of GRB 050401 Afterglow :
A case for Double Jet Model}

\author[Kamble et al.] 
       {Atish Kamble$^{1,2}$\thanks{e-mail : akamble@science.uva.nl}, Kuntal Misra$^{3,4}$, D. Bhattacharya$^{1,4}$, Ram Sagar$^{3}$\\ 
        $^{1}$ Raman Research Institute, Bangalore - 560 080\\
	$^{2}$ Astronomical Institute ``Anton Pannekoek", Kruislaan 403,
	1098 SJ Amsterdam, The Netherlands\\
        $^{3}$ Aryabhatta Research Institute of Observational Sciences, Manora Peak, Nainital - 263 129\\
	$^{4}$ Inter University Centre for Astronomy and Astrophysics, Post Bag 4, Ganeshkhind, Pune - 411 007}
\date{Received ---; accepted ---}
\maketitle
\label{firstpage}
%-------------------------------------------ABSTRACT------------------------------------%
\begin{abstract}

The afterglow of GRB 050401 presents several novel and interesting features :
\begin{enumerate}
	\item An initially faster decay in optical band than in X-rays.
	\item A break in the X-ray light curve after $\sim$ 0.06 day with
		an unusual slope after the break.
	\item The X-ray afterglow does not show any spectral evolution across the break 
		while the R band light curve does not show any break.
\end{enumerate}

We have modeled the observed multi-band evolution of the afterglow of GRB 050401 
as originating in a two component jet,
interpreting the break in X-ray light curve as due to lateral expansion of a
narrow collimated outflow which dominates the X-ray emission. The optical emission
is attributed to a wider jet component. Our model reproduces
all the observed features of multi-band afterglow of GRB 050401.

We present optical observations of GRB 050401 using the 104-cm Sampurnanand Telescope 
at ARIES, Nainital. Results of the analysis of multi-band data are presented and 
compared with GRB 030329, the first reported case of double jet.\\

\end{abstract}

%-------------------------------------------KEYWORDS---------------------------------------%

%------------------------------------------INTRODUCTION-------------------------------------%
\section{Introduction}
\label{sec:Introduction}
The optical and X-ray light curves of Gamma Ray Burst (GRB) afterglows, 
in the simplest cases, show a power law decay with an index $\alpha \sim 1.0$. 
Deceleration of the relativistic shock wave
generated by the explosion which results in GRB can explain the power law decay
of the GRB afterglows. The most common deviation from the power law decay 
behaviour of the afterglow light curves is an achromatic break seen in 
the light curve. 
This break has been seen in a significant number of GRB afterglows 
and has been successfully explained as being due to the sideways expansion 
of the collimated ejecta from the explosion.
In the post \emph{Swift} era, many more deviations from this simple behaviour 
of the afterglow light curve have been detected. \emph{Swift} with its 
capabilities of quick slewing towards the source has been able to observe
GRB afterglows as early as a few tens of seconds after the burst. 
In this early part of the evolution 
the GRB afterglows commonly exhibit a steep decay with $\alpha \sim$ 3 to 5 
with the usual definition $F_{\nu}(t) \propto t^{-\alpha}\nu^{-\beta}$ 
where $F_{\nu}(t)$ is the observed afterglow flux at frequency `$\nu$' and 
time $t$. The phase of steep decay lasts for about a few hundred seconds 
after which a slower decay, with $\alpha \sim 0.5$, of the afterglow starts.
About a few thousands of seconds after the burst the afterglow starts 
decaying steeply again with $\alpha > 1.0$.

Many GRB afterglows observed by \emph{Swift} show puzzling features
in the light curves like (1.) early steep decay [$\alpha \sim 3 ~to~ 5$] and
(2.) Chromatic breaks (breaks seen in some wavebands but not others) with $\Delta \alpha \sim 1.0$
which are difficult to explain using the standard fireball model \citep{Rees1992,Meszaros1993}.
It has been shown by \citet{PTO2006,RW2006} that the puzzling features of the X-ray afterglow 
light curves can be fitted using one or two components with exactly 
the same empirical functional form, 
viz. an exponential fall followed by a powerlaw decay of flux with time,
although it has not yet resulted into any physical understanding 
of the behaviour of the X-ray afterglow. 
While there is no clear understanding of the early steep decays of GRB afterglows, 
a few plausible explanations have been put forward : see e.g. \citet{Steep01,Steep02}. 
The flat decay of X-ray afterglow light curves which follows the steep decay
have been, in some cases, explained as being due to energy
injection from the central engine, probably a magnetar \citep{Zhang_01,Zhang_02}.
From the study of chromatic breaks seen in six well sampled afterglow
light curves \citet{Pan2006} concludes that if both, the optical
and the X-ray afterglows, were to arise from the same outflow
then the chromaticity of light curve breaks can rule out energy injection
or the structure of the jet as the possible reasons of it.

One such GRB afterglow with puzzling features in optical and X-ray light curves
is GRB 050401.
GRB 050401 triggered \emph{Swift}-BAT at 14:20:15 UT on 2005 April 01 \citep{GCN3162}. 
The X-ray afterglow was detected by \emph{Swift}-XRT \citep{GCN3161} about 130 seconds
after the trigger and the optical afterglow candidate was confirmed by 
ground based observations by \citet{GCN3164}. 
The burst duration $\rm T_{90}$ is estimated to be $\sim$ 33 seconds \citep{GCN3173}.
Using the measured spectral redshift of the afterglow ($z = 2.9$) \citep{GCN3176}
and the fluence \citep{GCN3173,050401_MDP,GCN3179} the isotropic equivalent energy
released during the explosion turns out to be  $1.4 \times 10^{54}$ for a flat universe 
with $\rm \Omega_{m} = 0.3, \Omega_{\Lambda} = 0.7 ~and~ H_{0} = 70 ~km~s^{-1} ~Mpc^{-1}$.\\

Multiband afterglow of GRB 050401 also presents some puzzling features which can be 
summarized as follows :
\begin{enumerate}
	\item A break in the X-ray light curve after $\sim$ 0.06 day with
		an unusual slope after the break \citep{050401_MDP,050401_Watson}.
	\item The X-ray afterglow does not show any spectral evolution across the break
		while the R band light curve does not show any break 
		\citep{050401_MDP,050401_Watson}.
	\item A large extinction inferred from X-ray afterglow which is not consistent
	      with the observed optical afterglow \citep{050401_Watson}.
\end{enumerate}

The optical observations are presented in \S~\ref{sec:Observations}. We have done 
some preliminary analysis of the light curves which is discussed in \S~\ref{sec:lightcurve}. 
We have tried to explain the multi-band behaviour of the GRB afterglow using 
a double jet model which is described in \S~\ref{sec:model} along with the previous 
attempts by others using a different model.
In the Discussion section (\S~\ref{sec:Discussion}) molecular clouds as a plausible explanation 
for the large extinction is presented (\S~\ref{sec:disc_extinction}).
The only other GRB afterglow which has been explained using a similar double jet model
is the GRB 030329 \citep{030329_38,030329_05}. We compare the physical features 
of GRB 030329 and GRB 050401 
in \S~\ref{sec:disc_compare}. Our conclusions are summarized in \S~\ref{sec:Summary}.

%------------------------OBSERVATIONS AND DATA REDUCTION--------------------------------------%
\section{Optical Observations and Data Reduction}
\label{sec:Observations}
Optical observations of the afterglow of GRB 050401 were carried
out in the broad and Johnson V and Cousins RI filters using the 104-cm Sampurnanand Telescope
of ARIES, Nainital on 01 April 2005.
The gain and read out noise of the CCD camera are 10 e$^-$/ADU and 5.3 e$^-$ respectively.
The data have been binned in 2$\times$2 pixel$^2$ to improve the signal-to-noise ratio.
The bias subtracted, flat fielded and cosmic ray removed images were processed
and analysed using
MIDAS \footnote{MIDAS is distributed by the European Southern Observatories.
Visit : www.eso.org/esomidas/}, 
IRAF \footnote{IRAF is distributed by the National Optical Astronomy 
Observatories, USA. Visit : http://iraf.noao.edu/} 
and DAOPHOT \citep{Stetson} softwares.

The \citet{Landolt1992} standard region SA 107 and the OA field in BVRI filters was observed 
on 16 May 2005 for photometric calibration during good photometric sky conditions. 
The values of atmospheric extinction on the night of 16/17 May 2005 determined from 
the observations of SA 107 bright stars are 0.26, 0.18, 0.13 and 0.10 magnitude 
in $B, V, R$ and $I$ filters respectively. The 7 standard stars in the SA 107 region cover a range 
of 0.339 $< (V-R) <$ 0.923 in color and 12.116 $< V <$ 14.884 in brightness. 

Using these transformation coefficients we determine BVRI magnitudes of 
18 secondary stars in GRB 050401 field and their average values are listed 
in Table~\ref{tab:calib}. 
The (X, Y) CCD pixel coordinates were converted to $\alpha_{2000}$, $\delta_{2000}$ values 
using the astrometric positions given by \citet{GCN3454}. The 18 secondary stars in the field 
of GRB 050401 were observed 2 to 4 times in $B, V, R$ and $I$ filters. 
These stars have internal photometric accuracy better than 0.01 mag. The zero-point 
differences on comparison between our photometry and that of 
\citet{GCN3454} are 0.15 $\pm$ 0.08, 0.09 $\pm$ 0.04, 0.10 $\pm$ 0.05 and 0.54 $\pm$ 0.29 
magnitude in $B, V, R$ and $I$ filters respectively. These differences are based on 
the comparison of the 6 secondary stars in the GRB 050401 field. 

The afterglow magnitudes were differentially calibrated with respect to the secondary stars 
listed in Table~\ref{tab:calib}. The magnitudes derived in this way are given 
in Table~\ref{tab:obs}.
%-----------------------------------TABLE 2 (050401 CALIBRATION STARS)-------------------------%
\begin{table}
\medskip
\begin{center}
\begin{tabular}{ccccccc} \hline\hline
ID&  $\alpha_{2000}$& $\delta_{2000}$ & $V$ & $B-V$ & $V-R$ & $V-I$\\
&(h m s)& (deg m s)& (mag) & (mag)&(mag)& (mag)\\ \hline
&&&&&&\\
1 &  16 31 20.01&  02 06 52.9&  17.28& 0.64& 0.35& 0.84\\
2 &  16 31 23.84&  02 07 44.3&  16.87& 0.77& 0.52& 0.98\\
3 &  16 31 29.22&  02 08 13.8&  17.66& 1.14& 0.75& 1.40\\
4 &  16 31 37.63&  02 08 07.3&  16.78& 0.69& 0.46& 0.87\\
5 &  16 31 40.12&  02 10 30.1&  16.35& 0.58& 0.38& 0.73\\
6 &  16 31 36.96&  02 11 36.5&  18.23& 0.97& 0.62& 1.12\\
7 &  16 31 32.61&  02 12 38.7&  17.68& 0.41& 0.34& 0.67\\
8 &  16 31 24.79&  02 13 35.4&  19.56& 1.37& 1.08& 2.29\\
9 &  16 31 18.94&  02 13 12.1&  19.21& 1.26& 0.83& 1.60\\
10&  16 31 18.56&  02 12 40.8&  15.30& 0.85& 0.51& 0.93\\
11&  16 31 22.46&  02 11 13.7&  15.61& 0.69& 0.46& 0.87\\
12&  16 31 21.38&  02 10 43.0&  15.51& 0.88& 0.53& 0.99\\
13&  16 31 19.42&  02 09 56.3&  14.60& 0.55& 0.36& 0.69\\
14&  16 31 15.08&  02 09 19.1&  14.34& 0.61& 0.38& 0.74\\
15&  16 31 23.42&  02 09 13.7&  16.39& 0.66& 0.44& 0.83\\
16&  16 31 17.26&  02 07 58.9&  16.17& 0.63& 0.41& 0.81\\
17&  16 31 15.93&  02 07 36.6&  18.91& 1.16& 0.89& 1.75\\
18&  16 31 14.79&  02 07 14.6&  17.54& 0.74& 0.47& 0.93\\
&&&&&&\\
\hline
\end{tabular}
\end{center}
\caption{The identification number (ID), ($\alpha$, $\delta$) for epoch 2000,
standard $V, (B-V), (V-R)$ and $(V-I)$ photometric magnitudes of the stars in the
GRB 050401 region are given.}\label{tab:calib} 
\end{table}
%------------------------------------------TABLE 4 (050401)---------------------------------%
\begin{table}
\medskip
\begin{center}
\begin{tabular}{cccc}\hline
\hline
Date (UT)	& $\Delta T$	&Magnitude	&Passband\\
2005 April	&	(days)	&(mag)		&\\
\hline
&&&\\
01.8824		& 0.2850	&22.33 $\pm$ 0.347&	V\\
&&&\\
01.8324		& 0.2850	&21.43 $\pm$ 0.231&	R\\
&&&\\	
01.8698		& 0.2724 	&20.51 $\pm$ 0.207&	I\\
&&&\\
\hline
\end{tabular}
\end{center}
\caption{The optical observations of the afterglow of GRB 050401 using the 104-cm 
Sampurnanand Telescope at ARIES, Nainital. $\Delta T$ in column 2 refers to the time
after the burst in days. The effective exposure time after combining all the images
turns out to be 900 $s$ for individual passbands reported here.}\label{tab:obs}
\end{table} 
%-------------------------------------------------------------------------%

\section{Light curves of GRB 050401 afterglow}
\label{sec:lightcurve}

Along with our own observations we have used observations reported 
elsewhere to study the light curves of GRB 050401. The X-ray light curve 
of GRB 050401 was obtained from \citet{050401_Watson}. 
The optical observations by \citet{050401_Watson} have been calibrated
by observing a Landolt field. We do not have a detailed information
about this calibration. Hence, to take into account any uncertainties 
associated with it, we have added an error of 0.2 magnitudes 
in the optical observations reported by \citet{050401_Watson}.
Another set of optical observations is taken from \citet{050401_ESR}
the calibration of which is roughly equivalent to the $R_{c}$ band system.
We add a small error of magnitude 0.1 to all these observations 
by ROTSE-III to take into account the calibration uncertainties.

VLA reported a $4 \sigma$ 
detection of a source at the position of GRB 050401 \citep{GCN3187} with intensity 
of 122 $\rm \mu Jy$ at 8.46 GHz about 5.7 days after the burst. Other attempts, 
including by GMRT in India at 610 MHz \citep{GCN3178} and by ATCA in Australia 
at 8.5 GHz and 4.8 GHz \citep{GCN3177}, to observe the radio afterglow 
of GRB 050401 could produce only upper limits. 

To construct the optical light curve we have corrected the observed magnitudes 
for the standard Galactic extinction law given by \citet{Mathis}.  The Galactic 
extinction in the direction of GRB 050401 is estimated to be E(B-V) = 0.065 
mag from the smoothed reddening map provided by \citet{Schlegel}.
The effective wavelength and normalisation given by \citet{Bessel} were used to 
convert the magnitudes to fluxes in $\mu$Jy. 

Most of the GRB afterglow light curves are well characterised
by a broken power law of the form 
\begin{equation}
F = F_{0}\{(t/t_{b})^{\alpha_{1}s} + (t/t_{b})^{\alpha_{2}s}\}^{-1/s}
\label{eqn:broken_powr_law}
\end{equation}
where $\alpha_{1}$ and $\alpha_{2}$ are the afterglow flux decay indices
before and after the break time ($\rm t_{b}$), respectively.
$\rm F_{0}$ is the flux normalisation and `s' is a smoothening parameter which
controls the sharpness of the break. Most known GRB afterglows have $\alpha_{1} \sim 1$
and $\alpha_{2} > \alpha_{1}$ i.e. the decay becomes steeper
after the break.\\

The X-ray and optical (R band) afterglow of GRB 050401 is very well sampled 
over a wide period of observation. The available $R$ band observations cover a duration 
from 36 s to 13 days after the burst while the X-ray observations range from $\sim$ 130 s to 12 days 
after the burst. The X-ray afterglow light curve shows a prominent break near 0.06 day while
the optical afterglow does not show any such break in the light curve.
We analyse this behaviour in detail below.\\
 
\begin{itemize}

\item[1.] The X-ray light curve shows a clear break near 0.06 day. 
The change of slope across the break is significant. 
Fitting Equation~\ref{eqn:broken_powr_law} to the data yield the decay slopes

$\rm \alpha_{X1} = 0.58 \pm 0.02$ for $\rm \Delta t < 0.06$ day;

$\rm \alpha_{X2} = 1.37  \pm 0.03$ for $\rm \Delta t > 0.06$ day;\\

The change of slope across the break, $\rm \Delta \alpha_{X} \sim 0.8$, is therefore 
quite substantial.

\item[2.] The optical afterglow shows a monotonic decay with decay index 
$\rm \alpha_{R} = 0.82 \pm 0.02$ over the entire period of observation (up to 13 days).
There is no evidence of a break simultaneous with that in the X-ray light curve.

\item[3.] According to the standard fireball model of GRB afterglows 
the X-ray light curve is expected to decay at least as fast as 
the optical light curve which is indeed true
for the majority of GRB afterglows observed so far. 
In the case of GRB 050401, we find that the X-ray afterglow shows a decay slower 
than optical light curve till $\sim$ 0.06 day after which it decays 
at a much faster rate as described above.

Thus, the relatively slow initial decay of optical and X-ray light curves, presence of a break 
in X-ray light curve and absence of such a break in optical,
and initially slower decay of the light curve in X-rays than in optical bands 
makes the afterglow of GRB 050401 an unusual and interesting one.

\end{itemize}

\section{Modeling of GRB 050401 afterglow}
\label{sec:model}

The change in slope across the break in the X-ray light curve  
$\rm \Delta \alpha_{X} \sim 0.8$ is too large to be explained by the passage 
of a spectral break.
In the standard fireball model of GRB afterglows the passage of the cooling break 
$\rm \nu_{c}$ through the observing band leads to a steepening of light curve 
by an amount $\rm \Delta \alpha_{X} = 0.25$, much smaller than 
that is observed for GRB 050401 afterglow, along 
with the change of spectral slope by $\rm \Delta \beta_{X} = 0.5$. 
The X-ray spectrum of GRB 050401 does not exhibit any change in the spectral 
slope across the break. We thus rule out the possibility of $\nu_{c}$ passing 
through the X-ray band at the time of break.

\citet{050401_MDP} explains the initial flatter decay 
and the break in the X-ray light curve based on a model by \citet{Zhang_01, Zhang_02}.
According to this model, the central engine of GRB remains active 
for several thousand seconds after the burst, continuously injecting energy into the fireball.
If the central engine is injecting energy above a certain critical rate then it can 
slow down deceleration of the shock wave which results in a shallow decay of the light curve.
The break in the light curve occurs when the central engine stops
injecting sufficient amount of energy into the fireball.
After this epoch the afterglow can be described using standard fireball
model giving $\alpha=(3/2)\beta$. Being a dynamical effect, the end of energy 
injection episode would result in an achromatic break in
the afterglow light curves. Although this model seem to explain 
the X-ray light curve reasonably well, absence of a similar break in optical 
afterglow light curve is sufficient to rule this model out for GRB 050401.

\citet{050401_Watson} point out another puzzle : the soft X-ray absorption implies 
an equivalent optical extinction of magnitude $\rm A_{v} = 9.1^{+1.4}_{-1.5}$
magnitudes in the host galaxy, assuming solar abundance.

However, if the optical and the X-ray emission are part of the same synchrotron spectrum, then 
$\rm A_{v}$ is constrained to be $\sim 1.45$ for no spectral break between optical and X-rays
and $\rm A_{v} < 0.67$ if a cooling break exists in between (an SMC extinction law is assumed).
These values are highly discordant with that predicted from X-ray absorption.
\citet{050401_Watson} suggests that this may indicate 
a non-universal dust to metals ratio which they estimate to be
more than a factor of 10 less than that in the SMC.

\citet{050401_Watson} remark that the only alternative to this highly anomalous
dust to metal ratio is separate emission regions for the optical and X-rays.
We explore this possibility assuming that two distinct jet components
give rise to the observed emission in these two (X-ray and optical) wavelength 
bands.  The jet contributing to the X-ray emission is narrow, exhibiting an early
break while that contributing to the optical emission is wider.  The optical
contribution from the narrow jet is strongly diminished due to the presence
of high extinction $\rm A_{v} \sim 9$ along the line of sight, while the wider
jet suffers from a smaller degree of average extinction.

\subsection{\bf Spectral Parameters of the Afterglow of GRB 050401}

The radiation spectrum of a GRB afterglows exhibits a power law spectrum
characterised by three break frequencies - the self absorption
frequency $\rm \nu_{a}$, the peak frequency $\nu_{\rm m}$ 
corresponding to the lower cutoff ($\gamma_{m}$) in the electron energy distribution
($n(\gamma_{e}) \propto \gamma_{e}^{-p}$, $\gamma_{e} > \gamma_{m}$ 
where $\gamma_{e}$ is the Lorentz factor of the radiating electrons) and 
the synchrotron cooling frequency $\nu_{\rm c}$.  The flux $F_{\rm m}$
at $\nu_{\rm m}$ provides the normalisation of the spectrum
\citep{Sari01}.

The photon index ($\Gamma$) of the X-ray afterglow is related 
to its spectral index ($\beta$), $\Gamma - 1 = \beta$,
which in turn is related to the electron energy
distribution index $p$ in any given spectral regime 
($\beta = p/2$ if $\rm \nu_{c} < \nu_{X}$ and 
$\beta = (p - 1)/2$ if $\rm \nu_{X} < \nu_{c}$).
The corresponding temporal decay index $\alpha_{\rm X}$ would be $(3p-2)/4$
and $3(p-1)/4$ respectively before the jet break and $p$ in both spectral
regimes after the jet break, according to the standard fireball model
for an afterglow expanding in a homogeneous interstellar medium.
In the present case, the observed values of $\alpha_{\rm X}$ 
are consistent with $p = 1.42$ and a jet break around $0.06$ days 
after the burst. However, we note that the observed value of 
the spectral photon index $\Gamma \sim 1.85 \pm 0.03$ \citep{050401_Watson}
implies a steeper $p \sim 1.7 \pm 0.06$. It should also be noted that
from analysis of the same data set of X-ray observations,
\citet{050401_MDP} infer $\beta = 0.75 \pm 0.15$ for PC mode data
after the break at 0.06 days. This $\beta$ is consistent with $p = 1.42$ that 
we inferred above.

The optical (R-band) afterglow, on the other hand, exhibits a temporal slope
$\alpha_{\rm R} = 0.82$ which, in the commonly encountered spectral regime
of  $\nu_{\rm m} < \nu_{\rm R} < \nu_{\rm c}$, implies $p=2.1$.  This is
different from that inferred for the X-ray afterglow, and indeed regardless
of the choice of spectral regimes it is not possible to produce both 
$\alpha_{\rm X}$ and $\alpha_{\rm R}$ from the same underlying
power-law energy distribution of injected electrons. 
One possibility, therefore, is that the optical and the X-ray afterglows 
originate in physically distinct outflows. We consider two physically 
distinct components of the outflow, such as the co-axial jets, one having 
a dominant contribution in the optical and the other in the X-rays, 
giving rise to the observed afterglow of GRB 050401.

\begin{figure*}
\begin{center}
\includegraphics[width=8.8cm]{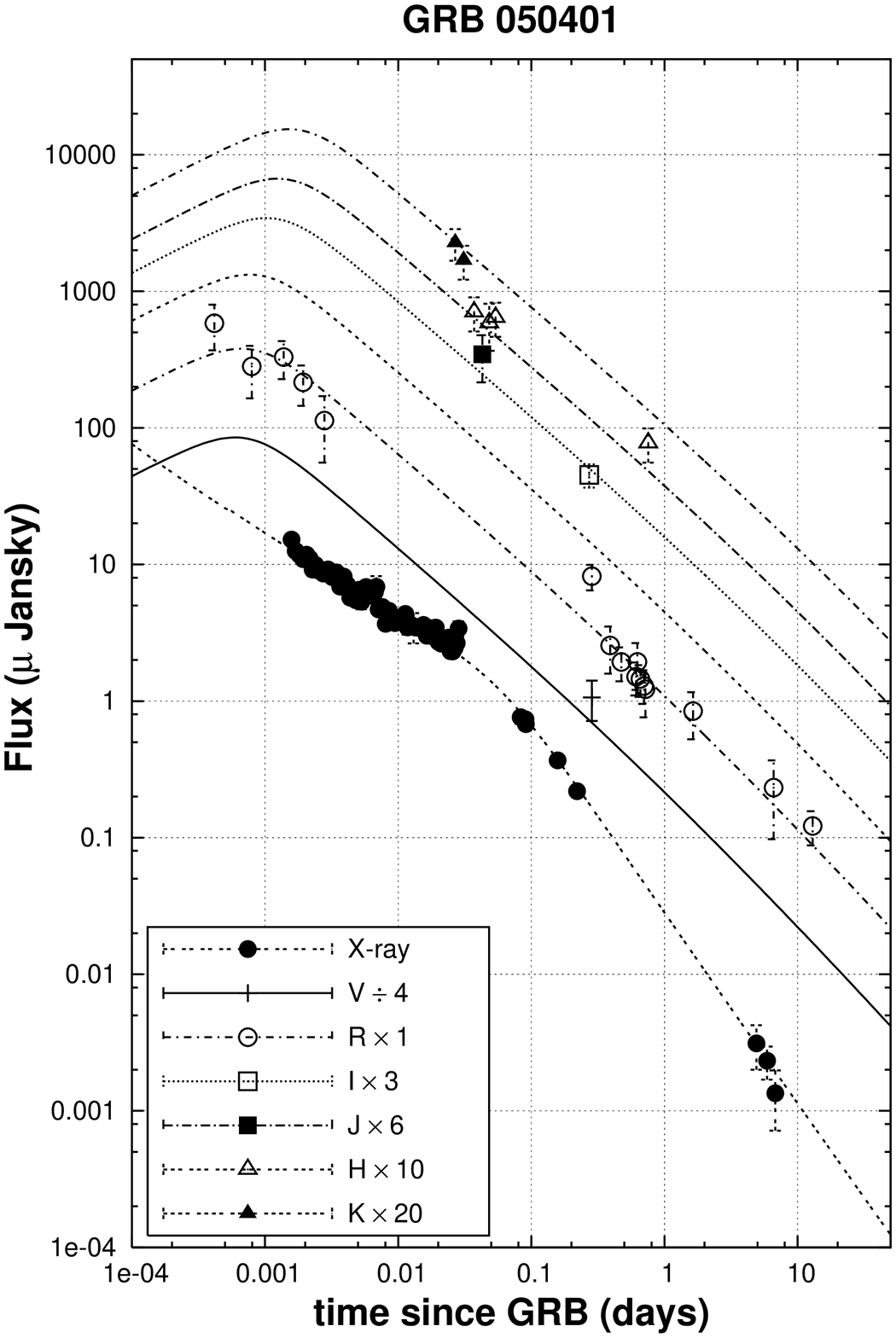}
\includegraphics[width=8.0cm]{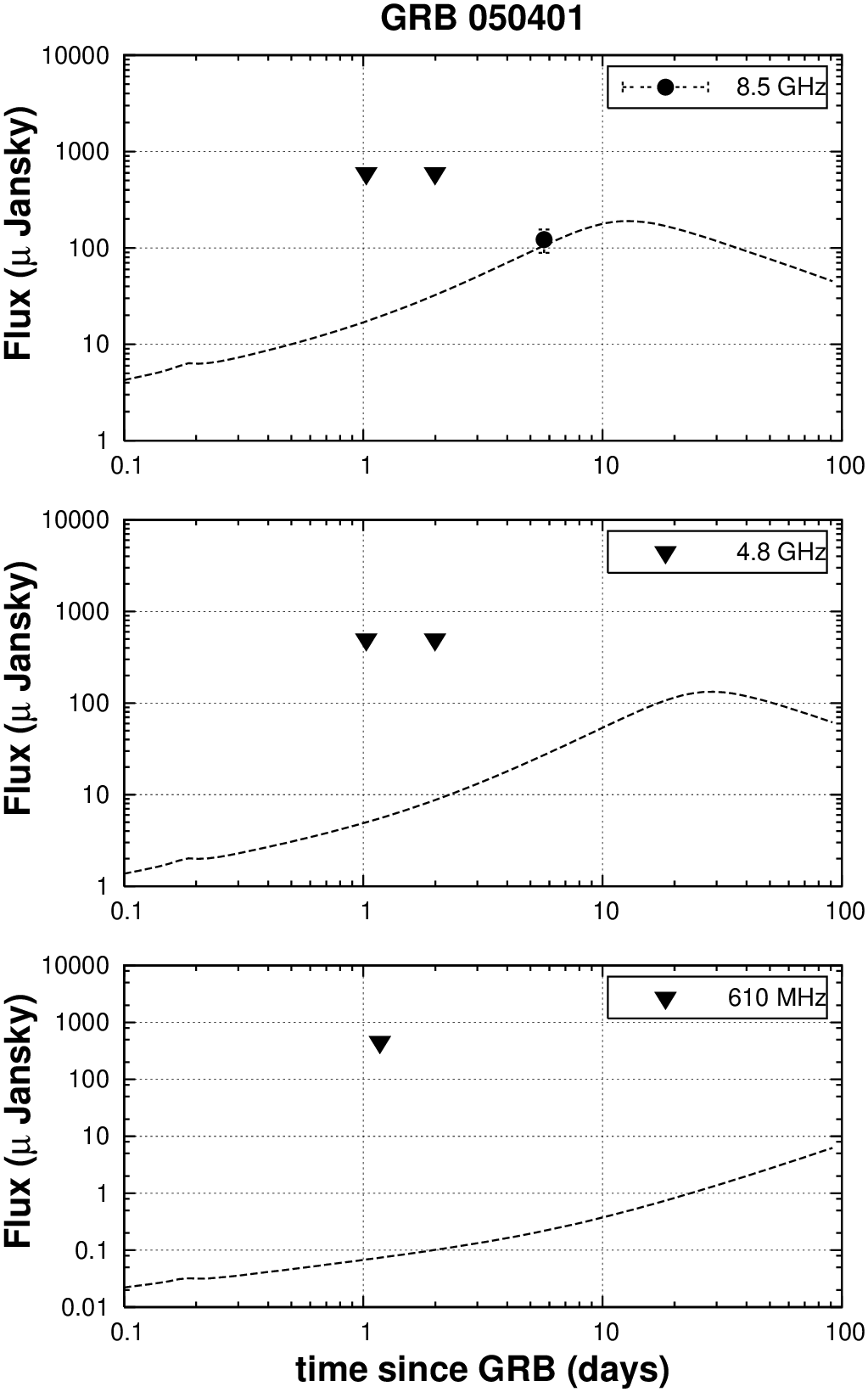}
\caption{The observed optical \& X-ray light curves (\emph{left panel}) 
and radio light curves (\emph{right panel})of GRB 050401 afterglow compared with 
the double jet model fit (solid lines). 
The steepening of X-ray afterglow light curves at 0.06 day 
after the burst is explained as a jet break due to the lateral expansion
of a narrow jet which has a dominant contribution in X-rays. 
The surrounding wider jet contributes dominantly in optical. Since, no break in 
the optical light curves is observed till 13 days after the burst, 
the  wider jet is expected to be $> 29^\circ$.
Our best fit model gives the value of electron energy distribution index 
within the narrow jet to be $p = 1.42$ and that within the wider jet to be $p = 2.1$.
The peak in the optical light curves corresponds to the passage of $\nu_{m}$
through the observing band.
The radio upper limits are indicated by filled triangles in the right panels.
The sole radio detection at 8.5 GHz is indicated by a filled circle. 
The solid lines in the right panels are the light curves expected from 
the best fit spectral parameters. The corresponding frequencies are listed
in a rectangle at the top right corner of each box.}
\label{fig:lightcurve}
\end{center}
\end{figure*}

We then fit the full, multi-band light curves of GRB050401 with those 
predicted by a double jet model using linear least square method.  Results
of this fit are displayed in Figure~\ref{fig:lightcurve}, and 
the best fit values of various spectral parameters are listed 
in Table~\ref{tab:spec_para}. We note that the contribution 
of the narrow jet to the optical afterglow is strongly suppressed due 
to large extinction. The X-ray afterglow, on the other hand, is modified
as a sum of the emission from both the jets, with the narrow jet being 
the dominant contributor. For the narrow jet we find a best fit value of
$p = 1.42$.  For the wider jet, which dominates the optical afterglow 
of GRB 050401, we estimate $p = 2.1$. The extinction that the radiation
from the narrow jet encounters is fixed at $A_{\rm v}$ = 9.1 
as derived from the soft X-ray absorption \citep{050401_Watson}, 
while that for the wide jet is treated as a fit parameter.

%-----------------------------------------------------------------------%
\begin{table}
\begin{center}
\begin{tabular}{|c|c|c|}
\hline
\hline
			&      Narrow Jet 	&     Wider Jet 	\\
\hline
$\rm \nu_{m} (Hz)$	& $2.0^{+1.2}_{-0.81}\times 10^{13}$	& $1.1^{+1.53}_{-0.83} \times 10^{13}$ 	\\
$\rm \nu_{c} (Hz)$	& $4.1 \pm 0.9 \times 10^{14}$		& $5.25^{+30.0}_{-5.0} \times 10^{15}$ 		\\
$\rm F_{peak}(\mu Jy)$	& $2140^{+210}_{-230}$                 	& $1750^{+1050}_{-950}            $      \\
$\rm t_{jet} (day)$	& $0.06 \pm 0.03$                 	& --                    		\\
$\rm p$			& $1.42 \pm 0.02$                	& $2.1^{+0.2}_{-0.11}$                  \\
$\rm E(B-V)_{Host}$	& 4.1            			& $0.23^{+0.22}_{-0.13}          $ 	\\
\hline
$\rm \chi^{2}_{dof} (dof)$&    \multicolumn{2}{c}{1.2 (85)}       \\
\hline
\end{tabular}
\end{center}
\caption{Best fit Spectral parameters for the afterglow of GRB 050401 
assuming two component jet model described in \S~\ref{sec:model}. Light curves generated using 
these parameters and their subsequent evolution according to the standard fireball model
are plotted in Figure~\ref{fig:lightcurve}. 
All the parameters are calculated at 0.01 day after the burst.
GRB 050401 was at redshift $z = 2.9$}\label{tab:spec_para}
\end{table}
%-----------------------------------------------------------------------%

\subsection{\bf Physical Parameters for GRB 050401}
Four spectral parameters ($\rm \nu_{a} ,
\nu_{m} , \nu_{c}\ and\  F_{peak}$) are related to four physical parameters
viz. n (number density of the circumburst medium), 
E (total energy content of the fireball), energy fraction in relativistic
electrons $\rm \epsilon_{e}$ and that in magnetic field $\epsilon_{B}$.
The typical value of self absorption frequency $\rm \nu_{a}$ lies in
radio-mm waves and hence is best estimated only if the afterglow is well
observed in these bands. Unfortunately, the afterglow of GRB 050401 
was detected only once at the radio band \citep{GCN3187} which is not 
sufficient to determine $\rm \nu_{a}$ accurately. 
We therefore converted the three remaining
spectral parameters into the four physical parameters using 
$\rm \epsilon_{e} = \sqrt{\epsilon_{B}}$ as an additional constraint.
The choice of this relation is motivated by \citet{Medvedev}.
When $p < 2.0$, as it is in the present case of narrow jet,
a high energy cut-off for the electron energy distribution
is required and the expressions for spectral parameters, as given 
in \citet{Wijers1999}, have to be modified accordingly. The modifications 
have been provided by \citet{DB2001} which we have used for estimating 
the physical parameters in the present case.
We estimate the density of the circumburst medium to be
$n \approx 10$ and $\rm \epsilon_{e} = \sqrt{\epsilon_{B}} = 0.03$
for both the jets. The physical parameters estimated for both the jets 
are listed in Table~\ref{tab:phys_para}.
Using the $E^{iso}$ and $n$, and the jet break time in X-rays,
$t_{j} = 0.06$ days, we find the opening angle of the narrow jet to be
quite small, $1.15^{\circ}$.
Since there is no jet break seen in the optical light curve till
$\sim 13$ days, a lower limit on the opening angle of the wider jet 
is derived to be $29^{\circ}$.
The collimation corrected kinetic energies are $E^{K}_{wide} > 6.5 \times 10^{50}$ ergs
and $E^{K}_{narrow} = 1.1 \times 10^{50}$ ergs.
%-----------------------------------------------------------------------%
\begin{table}
\begin{center}
\begin{tabular}{|c|c|c|}
\hline
\hline
		&      Narrow Jet 		&     Wider Jet 	    \\
\hline
$n$		& $14.7^{+10.5}_{-5.34}$		&  $20^{+2583}_{-19.3}$	    \\ 
$\epsilon_{e}$	& $(2.3 \pm 0.6)\times 10^{-2}$	&  $(4^{+6}_{-2})\times 10^{-2}$   \\
$\epsilon_{B}$	& $5^{+4}_{-2} \times 10^{-4}$&  $1^{+9}_{-0.9} \times 10^{-3}$\\ 
$E_{52}^{iso}$	& $53.23 \pm 16.2$       &  $1.34^{+1.36}_{-0.82}$   \\
$\theta_{j}  $  & $1.15^{\circ} \pm 0.15^{\circ}$& $> 29^{\circ}$           \\
$E_{52}^{corr}$	& $(1.1 \pm 0.2) \times 10^{50}$&  $> 6.5 \times 10^{50}$   \\
\hline
\end{tabular}
\end{center}
\caption{The physical parameters for the afterglow of GRB 050401 assuming 
a two component jet model described in \S~\ref{sec:model}. The quantity
$E_{52}^{iso}$ is the isotropic equivalent energy in units of $10^{52}$ ergs.
The corresponding collimation corrected energy is $E_{52}^{corr}$
in units of $10^{52}$ ergs. 
}\label{tab:phys_para}
\end{table}
%-----------------------------------------------------------------------%
%------------------------------------FIGURE 3 (CHART: 050401)-------------------------------------%
%-------------------------------------------------------------------------%
\section{Discussion}
\label{sec:Discussion}

\subsection{A plausible explanation for the large extinction inferred from X-ray absorption}
\label{sec:disc_extinction}

\begin{figure*}
\begin{center}
\includegraphics[width=6cm,angle=-90]{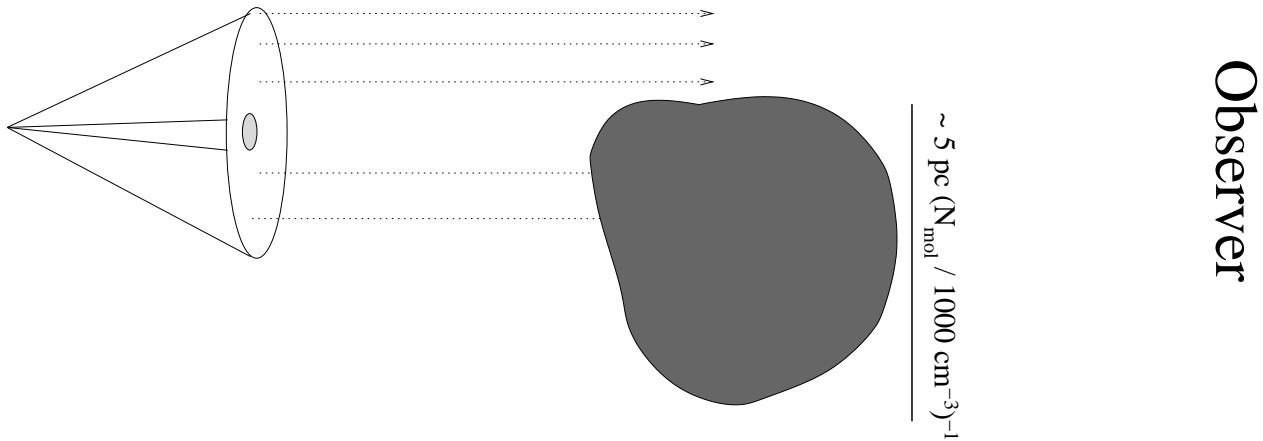}
\caption{A side view of the double jet (not to scale). 
The observer is on the axis of the jets
and at a distance of 24 Gpc (which can be considered as at infinity for geometric
purposes in this figure). The arrows indicate the afterglow light rays
emanating from the jets. 
The intervening molecular cloud, of size larger than 5 to 55 parsecs, 
responsible for the observed large extinction is sitting at a distance 
of about 100-1000 parsecs from the GRB.
The estimated diameters of the jets around 0.05 days
turn out to be about $2 \times 10^{-3} ~pc$ and $> 2 \times 10^{-2}~pc$ 
respectively for the narrow and the wide jet.
The large cloud covers a significant portion of the central narrow jet
and partially covers the wide jet when seen from the observer's point of view.
As a result, the optical radiation from the narrow jet is completely extincted.
Most of the optical radiation from the wide jet does not suffer from
this extinction.}
\label{fig:doublejet} 
\end{center}
\end{figure*}

It is now well established from the observations of GRB hosts that long GRBs preferentially
occur in massive star forming regions e.g. \citet{WB2006}. The massive star forming regions
host large molecular clouds. Typical column densities of cold molecular clouds 
are $\rm > 10^{22} ~cm^{-2}$, densities $100 - 10^{4} cm^{-3}$ 
and sizes $\rm \sim 20 ~pc$. Giant molecular clouds are even denser 
($\rm 10^{4} - 10^{7} cm^{-3}$) and larger ($\rm \sim 100 ~pc$) \citep{Shore}. 
It is possible that one such cloud in the host galaxy of GRB 050401 happens 
to fall along our line of sight which can explain the large extinction inferred 
from the X-ray spectrum. We consider the possibility of radiation from 
the double jet 
of GRB 050401 being obscured by a molecular cloud so aligned that it covers 
the narrow jet of GRB 050401 completely while the wide jet is partially covered. 
By changing the fractional coverage of wide jet by the cloud we measured 
change in the value of reduced $\chi^{2}$ of the fit.
In effect, this amounts to adjusting the intrinsic luminosity of the wide jet 
upwards with increasing covering factor to match the observed optical flux. 
This results in the relative contribution of the wide jet to the X-ray afterglow 
to increase, affecting the fit quality. Keeping all other parameters fixed 
at their best fit values obtained for zero coverage, we find that a covering 
fraction of 60\% can be accomodated within a range of $\Delta \chi^{2}/dof = 1$. 
Beyond this the reduced $\chi^{2}$
rises sharply and reaches $\Delta \chi^{2}/dof > 15$ for 
a covering factor of $\sim 90\%$. 
For the observed column density of 
$1.7 \times 10^{22}$ cm$^{-2}$ \citep{050401_MDP},
and assuming typical densities ($100-1000$ cm$^{-3}$) of the molecular clouds,
the size of the molecular cloud could be estimated to be around $5-55$ parsecs.
It is therefore probable 
that one such molecular cloud partially obscures our view of  
GRB 050401. This situation is illustrated in Figure~\ref{fig:doublejet}.
%-------------------------------------------------------------------------%
%-------------------------------------------------------------------------%
At this point, we would like to point out two possible 
caveats in the double jet model proposed here :

The separation of the optical and X-ray emitting regions, as 
proposed in the present model, is motivated by the large
discrepancy of  about 8 magnitudes between the amount of 
optical extinction inferred from soft X-ray absorption and 
that from observed optical-IR spectrum of the GRB050401 
afterglow.  It should, however, be kept in mind that 
the \citet{PS1995} relation used to predict 
$A_V$ from X-ray absorbing column $N_H$ is an empirical one, 
and cannot be considered fully reliable in all circumstances.
For example, a metallicity higher than solar by a factor of 
~10, or a dust-to-gas ratio lower by a similar factor, can
reconcile the X-ray absorption with observed optical 
extinction.  Such explanations in this case cannot be
ruled out, and have been already discussed by 
\citet{050401_Watson}. 

The second caveat is that the model presented here requires
a rather special geometrical alignment - the two jets of
the GRB should shine through the outer edge of a molecular 
cloud, much larger in size than the transverse extent of the
jet working surface, in such a manner as to provide large 
extinction to the inner jet but much less to at least half 
the outer component. This requires that the outer edge of 
the cloud be dense, and have a strong density gradient to
differentially affect the two jet components.  An elongated,
cigar-shaped cloud with its axis nearly parallel to the line 
of sight, would also help such a scenario. We also note that
the size of the cloud required, as estimated by us using an
average density, is prone to large uncertainties if its 
shape is unusual or if large density gradients are present.
%-------------------------------------------------------------------------%
\subsection{GRB 050401 and GRB 030329 : A comparison}
\label{sec:disc_compare}

The only other GRB whose afterglow has been explained as being due 
to double jet is the GRB 030329 \citep{030329_38, 030329_05}.
optical and X-ray light curves of GRB 030329 afterglow showed a near simultaneous 
break at 0.55 day whereas the radio light curves had a break at about 10 days 
after the burst. \citet{030329_38} have explained the two breaks as being due to 
lateral expansion of the two co-axial jets of different opening angles 
($\sim 5^{\circ}$ and $\sim 17^{\circ}$).

In the case of GRB 050401, afterglow light curves do not show the presence 
of two different breaks. Instead, absence of a break at optical frequencies
till late times ($\sim 13$ days after the burst)
leads us to infer the presence of a wider jet with opening angle
larger than $29^{\circ}$ while a steep break ($\Delta \alpha \sim 0.8$)
at 0.06 day after the burst in X-ray light curve can be explained 
as a jet break due to lateral expansion of a narrow jet of opening 
angle $1.15^{\circ}$.

The wider jet of GRB 030329 was estimated to be marginally more energetic 
than the narrower jet \citep{030329_38, 030329_05}. Similarly, 
in the case of GRB 050401, we find, that the wider jet is marginally 
more energetic than the narrower jet.

\subsection{GRB 050401 and the Ghirlanda Relation :}
\label{sec:Ghirlanda_relation}

It has been found that the collimation corrected energies ($E_{\gamma}$) 
of the GRBs are correlated with the peak energy of the GRB spectrum 
as measured in the frame of reference of the source ($E_{peak}^{src}$). 
This correlation is also called as
the Ghirlanda relation \citep{Ghirlanda2004}. Unfortunately, the $E_{peak}^{src}$
for GRB 050401 is not available as it falls outside the energy range
of BAT. However, \citet{050401_Sato} have used the Konus-Wind spectral 
data to find $E_{peak}^{obs}$. From their analysis \citet{050401_Sato}
finds that in order to satisfy the Ghirlanda relation the afterglow
light curve of GRB 050401 should exhibit a jet break $\sim 10^{4} ~s$ 
after the burst. This lower limit of the allowed range for jet break 
time is close to the break seen at 0.06 day in the X-ray light curve 
of GRB 050401, which we interpret as a jet break corresponding 
to the narrow jet in our model.

\citet{050401_Sato} quantifies the Ghirlanda relation 
as $E_{peak}^{src} = A ~E_{\gamma,52}^{0.706}$ where $E_{\gamma,52}$ 
is the collimation corrected energy released in $\gamma$ rays
during the burst, in units of $10^{52}$ ergs. Using a sample of 
a large number of GRBs \citet{050401_Sato} constrains the value 
of the proportionality constant $A$ : $1950 < A < 4380 $.
Using the estimated value of $E_{\gamma}^{iso} \sim 10^{54}$ ergs
and the $1.15^\circ$ as the opening angle of the narrow jet 
in our double jet model, the $E_{\gamma}$ turns out 
to be $2 \times 10^{50}$ ergs. Using $E_{peak}^{src} = 447^{+75}_{-64} keV$
for GRB 050401 as reported by \citet{050401_Sato} along with 
$E_{\gamma} = 2 \times 10^{50}$ ergs we estimate $A = 7076^{+2597}_{-1897}$. 
This value of $A$ is within $2~\sigma$
of $A = 4380$, the higher limit on $A$ obtained considering 
the sample of GRBs satisfying the Ghirlanda relation.
Having discussed this, we would also like to point out that 
the Ghirlanda relation has sometimes been critisized as being 
due to selection effects rather than being an intrinsic 
correlation \citep{Butler2008}.

\section{Summary}
\label{sec:Summary}

We have reported VRI band observations of GRB 050401 afterglow on 1st Apr. 2005.
Also, we have modeled the afterglow of GRB 050401 as due to two physically distinct
collimated outflows, using 
our own VRI band photometry along with the observations available 
in the literature, and compared with GRB 030329.
Our main conclusions about GRB 050401 are as follows :
\begin{itemize}
\item [1.] We showed that the light curves of GRB 050401 afterglow can not be explained
	under the assumption of continuous energy injection. The flatter decay, which appealed for 
	the continuous energy injection model, can instead be explained by low values 
	of electron energy distribution index $p$.
\item [2.] The afterglow of GRB 050401 can be well fit by the double jet model 
	with the interpretation that the break in the X-ray light curve at $\sim$ 0.06 day
	after the burst is due to a narrow collimated jet expanding sideways. 
	The obscured optical emission is attributed to a wider which did not
	undergo significant sideways expansion until 
	at least $\sim 13$ days after the burst.
\item [3.] Kinematically, we find that the wider jet is slightly more energetic,
	than the narrow jet. This result is similar to what was found in 
	the double jet of GRB 030329.
\item [4.] Our interpretation of the break in the X-ray light curve 
	at 0.06 days after the burst as a jet break is consistent 
	with the Ghirlanda relation.
\end{itemize}

\section*{Acknowledgements}
We are thankful to the anonymous referee for his/her detailed comments which
have improved the paper significantly. This research has made use of data 
obtained through the High Energy Astrophysics Science Archive Research 
Center Online Service, provided by the NASA/Goddard Space Flight Center. 
We acknowledge the use of public data from the \emph{Swift} data archive.
Two of the authors AK and KM acknowledge the support received from
Dept. of Science and Technology, Govt. of India.

\end{document}